\documentclass[epj]{svjour}
\usepackage{epsfig} \usepackage{psfig} \usepackage{times}

\begin{document}

\title{Forward-backward angle asymmetry and polarization observables in high-energy
deuteron photo\-disintegration}
\author{V.Yu.~Grishina $^a$, L.A.~Kondratyuk $^{b}$, W.~Cassing $^c$, E.~ De~Sanctis $^d$, M.~Mirazita $^d$,
F.~Ronchetti $^d$ and P.~Rossi $^d$ \\}
\institute{$^a$ Institute for Nuclear Research, 60th October Anniversary Prospect 7A, 117312 Moscow, Russia\\
$^b$ Institute for Theoretical and Experimental Physics, B.\
  Cheremushkinskaya 25, 117218 Moscow, Russia \\
$^c$ Institute for Theoretical Physics,  University of Giessen, Heinrich-Buff-Ring 16, D-35392 Giessen, Germany\\
$^d$ INFN-Laboratori Nationali di Frascati, CP 13, via E. Fermi, 40; I-00044, Frascati, Italy}
\date{Received: date / Revised version: date}

\abstract{Deuteron two-body photodisintegration is analysed within the framework of the Quark-Gluon Strings Model.
It is found that the forward-backward angle asymmetry predicted by the model is confirmed by the recent
data at different photon energies from Jlab. New calculations for polarization observables, the cross section
asymmetry $\Sigma$ and the polarization transfer $C_{z^{\prime}}$, for photon energies  $(1.2\div6)$~GeV are
presented and compared with the data available up to 2~GeV.}


\authorrunning{V. Yu. Grishina et al.}

\titlerunning{Polarization observables...}
\maketitle
\section{Introduction}

Experiments on high energy two-body photodisintegration of the deuteron \cite{Bochna,Schulte} have shown that the
cross section data at proton angles $\theta_p^{\mathrm{CM}} = 89^{\circ}$ and $69^{\circ}$ exhibit scaling
consistent with the
constituent quark counting rule behavior \cite{Matveev}\footnote{i.e. at fixed c.m.
angle the differential cross
section $d\sigma/dt_{\gamma d \to pn}$ scales as  $\sim s^{-11}$}
for photon energies $E_{\gamma} \geq~1$~GeV
while at forward angles, $\theta_p^{\mathrm{CM}} = 36^{\circ}$ and $52^{\circ}$,
scaling is not observed for
$E_{\gamma} \leq~4$~GeV.
Moreover, the data on the recoil polarization for the
$d(\overrightarrow{\gamma},\overrightarrow{p})n$ reaction at
$\theta_p^{\mathrm{CM}} = 90^{\circ}$ for photon energies up to
2.4~GeV \cite{Wijesooriya} do not support the
helicity conservation as predicted by perturbative QCD (pQCD). Thus scaling is no
longer considered as sufficient evidence for the
applicability of pQCD in the energy range  $E_{\gamma} = (1\div 4)$~GeV and
nonperturbative approaches should be applied, too.

To this aim, recently, some of us have studied the high-energy deuteron photodisintegration within the framework
of the Quark-Gluon Strings Model (QGSM) \cite{Grishina,Kondratyuk}. This  model - proposed by Kaidalov
\cite{Kaidalov,Kaidalov99} - is based on two ingredients: i) a topological expansion in QCD and ii) a space-time
picture of the interactions between hadrons that takes into account the confinement of quarks. In a more general
sense the QGSM can be consi\-dered as a microscopic (nonperturbative) model of Regge phenomenology for the
analy\-sis of exclusive and inclusive hadron-hadron and photon-hadron reactions on the quark level.

We recall that originally the QGSM has been formulated for small scattering angles or low 4-momentum transfer
(squared) $t$ (here $t$, $u$ and $s$ are the usual Mandelstam variables).
The question thus arises, how to
extrapolate the QGSM amplitudes to large angles (or large $t$). Following Coon et al. \cite{Coon} we assume that
there is only a single analytic Regge term that smoothly connects the small angle and fixed-large-angle regions.
Thus, according to Ref.\cite{Arik}, we require that the amplitude at fixed angle  should be obtained either as the
large $t$ limit of the forward Regge form or as the large $u$ limit of the backward Regge form. The only
solution to these boundary conditions is  a logarithmic decreasing trajectory,
\begin{equation}
\alpha_N (t)=\alpha_N(-T_B) - d\,  \mathrm{ln} (-t/T_B), \label{LDM}
\end{equation}
where $d$ is a constant and $T_B$ is a scale parameter. Such a form of the Regge trajectory naturally arises in
the logarithmic dual model (LDM), that very well describes the differential cross section $d\sigma/dt$ for elastic
 $pp$ scattering in the energy range $(5\div24)$~GeV for $-t$ up to 18 GeV${}^2$ \cite{Coon}. It is worth noti\-cing
that logarithmic form of non-linear Regge trajectories have also been discussed in
Refs.~\cite{Chikovani,Ito,Bugrij,Brisudova}.
The special case with $d=0$ corresponds to  'saturated' trajectories, which means
that all the trajectories
approach a constant at large negative $t$. We recall, that this case leads to the
constituent-interchange model which is a predecessor of the asymptotic
quark-counting rules \cite{Matveev,Brodsky}. The approach
with 'saturated' trajectories was successfully used to explain the
large $t$ behavior of hadron and photon induced reactions in
Refs. \cite{Guidal,Fiore,White,Battaglieri}.

Within the QGSM the deuteron photodisintegration amplitude $T(\gamma d \to pn)$ can be described in first
approximation by planar graphs with three valence quark exchange in $t$ (or $u$)-channels, which corresponds to a
nucleon Regge trajectory (see  Fig. \ref{fig:qgsm}). In Ref. \cite{Grishina} deuteron photodisintegration has been
analyzed using nonlinear Regge trajectories. It was found that the QGSM provides a reasonable description of the
Jlab data on deuteron photodisintegration at large momentum transfer $t$ \cite{Bochna} when using  a logarithmic
form for the nucleon trajectory similar to that of Eq. (\ref{LDM}).
This has provided new evidence for a nonlinearity of the Regge trajectory $\alpha_N(t)$.

In this work we compare the predictions of the QGSM with all data available at high energies
\cite{Bochna,Schulte,Crawford,Napolitano,Freedman,Beltz,Gilman,Mirazita,Ronchetti,Rossi}
and provide a more detailed analysis of the
forward-backward angle asymmetry. Moreover, as a novel aspect we calculate the cross section asymmetry $\Sigma$ and the
polarization transfer to the proton $C_{z^{\prime}}$ in the reaction
$\overrightarrow{\gamma}d \to \overrightarrow{p}n$ for photon energies  $E_\gamma = (1.2\div6)$~GeV and compare them to
the data available at $\theta_p^{\mathrm{CM}} = 90^{\circ}$ and
up to 2~GeV \cite{Wijesooriya,Adamian}.

The layout of the paper
is as follows: In Section 2 the spin structure of the $\gamma d \to p n$
amplitude is eva\-luated and in Section 3
the analysis of the forward-backward asymmetry is discussed. In Section 4 the results (and predictions) of the
QGSM are compared with the available data for the energy dependence of the differential cross section at fixed
angles. Section 5 is devoted to the definition and calculation of polarization observables while Section 6
summarises the results of the work.

\section{Spin structure of the $\gamma d \to p n$ amplitude}

\begin{figure}[h]
\centerline{\psfig{figure=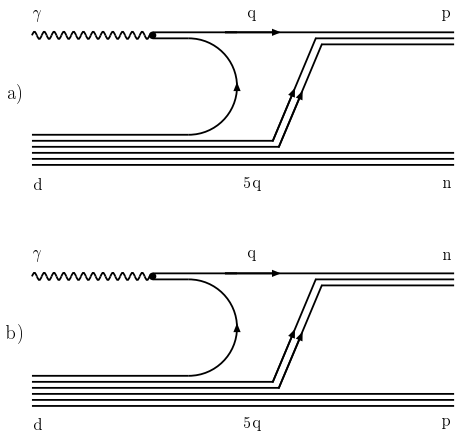,width=6.0cm}}
    \caption{Diagrams for deuteron photodisintegration describing
    three valence quark exchanges in the $t$- (a) and
      $u$-channel (b).}  \label{fig:qgsm}
\end{figure}

As already mentioned above the  main assumption of the QGSM is that the deuteron photodisintegration amplitude
$T(\gamma d \to pn)$ can be described  by planar graphs with three valence quark exchange in $t-$ or $u-$ channels
with any number of gluon exchanges between them (cf. Fig.\ref{fig:qgsm}). This corresponds to the contributions
of the $t$- and $u$- channel nucleon Regge trajectories. In the space-time picture the intermediate $s$-channel
consists of a $6q$ string (or color tube) with $q$ and $5q$ states at the ends.

The spin dependence of the  $\gamma p \to pn$ amplitude has been evaluated in Ref. \cite{Grishina} in a simple
way by assuming that all intermediate quark clusters have minimal spins and the $s$-channel helicities in the
quark-hadron and hadron-quark transition amplitudes are conserved. In this limit the spin structure of the
amplitude $T(\gamma d \to pn)$ can be written as (see Ref. \cite{Grishina}, comment after Eq.~(27))
\begin{eqnarray}
\lefteqn{ \langle p_3,\lambda_{p}; p_4,\lambda_{n} |
\hat{T}\left(s, t\right)|
p_2,\lambda_{d};p_1,\lambda_{\gamma}\rangle \simeq }\nonumber \\ &&\bar
u_{\lambda_p}(p_3) \hat {\epsilon}_{\lambda_{\gamma}}
\left[{A(s,t)( \hat{p}_3-\hat{p}_1) +B(s,t) m}\right] \hat
{\epsilon}_{\lambda_{d}} v_{\lambda_n}(p_4), \label{spin1}
\end{eqnarray}
where $m$ is the nucleon mass, $p_1$, $p_2$, $p_3$, and $p_4$ are the 4-momenta
of the photon, deuteron, proton,
and neutron, respectively, and $\lambda_i$ denotes the $s$ channel
helicity of the $i$-th particle.
The invariant amplitudes $A(s,t)$ and $B(s,t)$ have similar
Regge asymptotics (see below). It is possible to show
(cf. Ref. \cite{Grishina}) that at small scattering angles the ratio $R=A(s,t)/B(s,t)$ is
a smooth function of $t$
and can be considered as an effective constant that depends on the ratio of the nucleon mass to the constituent
quark mass $m_q$: $R\simeq m/(2 m_q)$. It is interesting to note that the spin
structure of the  $\gamma d \to pn$
amplitude in Eq. (\ref{spin1}) is very similar to the amplitude within the
Reggeized Nucleon Born Term Model
(see Refs. \cite{Guidal,Irving}) where the ratio $R=A/B=1$ is directly related to the nucleon propagator. In line with
Ref. \cite{Grishina} we also treat the  ratio $R$ as a
constant value that, however, may range between 1 and 2.5.

The differential cross section for the reaction $\gamma
d\to pn$ is
\begin{eqnarray}
&\displaystyle \frac{d\sigma _{\gamma d\to pn}}{d t}& =
\frac{1}{64\,\pi s}\ \frac{1}{(p_{\gamma \, \mathrm{cm}})^2}\
\nonumber\\
&&\times \left[ S_{t}\ \left|B^{(+)}(s,t)\right|^2+
S_u\ \left|B^{(-)}(s,u) \right|^2 \nonumber \right.\\
&& \left. + 2S_{tu}\ {\mathrm
{Re}}\, (B^{(+)}(s,t) B^{(-)*}(s,u))
\vphantom{\left|B^{(+)}(s,t)\right|^2}
\right], \label{eq:sigt}
\end{eqnarray}
where the amplitudes $B^{(\pm)}(s,t)=B^{(\rho)}(s,t)\pm B^{(\omega)}(s,t)$ are combinations of the contributions
from isovector ($\rho$ like) and isoscalar ($\omega$ like) photons. The kinematical functions $S_t$,
$S_u$, $S_{tu}$ in (\ref{eq:sigt}) can be written in a covariant form as
\begin{eqnarray}
S_t=\frac{1}{6}  \sum _{\lambda_{\gamma},\ \lambda_{d}}
{\mathrm {Sp}}\, \left[\vphantom{\hat {\epsilon}^*_{\lambda_{\gamma}}}
\hat {\epsilon}_{\lambda_{\gamma}}
\left(R \,( \hat{p}_3-\hat{p}_1)
+ m \right)
\hat {\epsilon}_{\lambda_{d}}
\left(\hat {p}_4- m  \right) \right.  \nonumber \\
\times \left. \hat {\epsilon}^*_{\lambda_{d}}
\left(R \, ( \hat{p}_3-\hat{p}_1)
+  m \right)
 \hat {\epsilon}^*_{\lambda_{\gamma}}
\left(\hat {p}_3+ m  \right) \right] \ ,\nonumber \\
S_u=\frac{1}{6}  \sum _{\lambda_{\gamma},\ \lambda_{d}}
{\mathrm {Sp}}\,  \left[ \vphantom{\hat {\epsilon}^*_{\lambda_{\gamma}}}
\hat {\epsilon}_{\lambda_{d}}
\left(R\, ( \hat{p}_4-\hat{p}_1)-  m \right)
 \hat {\epsilon}_{\lambda_{\gamma}}
\left(\hat {p}_4- m  \right) \right. \nonumber \\ \times \left.
\hat {\epsilon}^*_{\lambda_{\gamma}} \left(R\, (
\hat{p}_4-\hat{p}_1) -  m \right) \hat
{\epsilon}^*_{\lambda_{d}} \left(\hat {p}_3+ m
\right)\right] \ ,\nonumber\\
S_{tu}=-\frac{1}{6} \sum
_{\lambda_{\gamma},\ \lambda_{d}} {\mathrm {Sp}}\,
\left[ \vphantom{\hat {\epsilon}^*_{\lambda_{\gamma}}}
\hat{\epsilon}_{\lambda_{\gamma}} \left(R \, ( \hat{p}_3-\hat{p}_1) +
m \right) \hat {\epsilon}_{\lambda_{d}} \left(\hat {p}_4- m
\right)\right. \nonumber \\ \times \left. \hat
{\epsilon}^*_{\lambda_{\gamma}} \left(R\, (
\hat{p}_4-\hat{p}_1) - m \right) \hat
{\epsilon}^*_{\lambda_{d}} \left(\hat {p}_3+ m
\right)\right].
\end{eqnarray}
In order to achieve consistency of the differential cross section $d\sigma/dt$
with Regge asymptotics for large $s$ and fixed $t$, we use the
following parametrization of the amplitude $B^{(+)}(s,t)$
\begin{equation}
\left|B^{(+)}(s,t)\right|^{2}=\frac{1}{C_{0}\, p_{\gamma \, \mathrm{cm}}^2} \
\left|\mathcal{M}_{\mathrm {Regge}}(s,t)\right|^2\ ,
\label{Bst}
\end{equation}
where $C_{0}=(36\pm3)~$GeV$^2$ and
\begin{equation}
\mathcal{M}_{\mathrm{Regge}}(s,t)= F(t)
\left(\frac{s}{s_0}\right)^{\alpha_{N}(t)} \exp{\left[
      -i\ \frac{\pi}{2}\left(\alpha_{N}(t) -
        \frac{1}{2}\right)\right]}\  .
\label{eq:Mregge}
\end{equation}
Here $\alpha_N(t)$ is the trajectory of the nucleon Regge pole and
$s_0 =4~\mathrm{GeV}^2 \simeq m_d^2$ ($m_d$
denoting the mass of the deuteron). We take the dependence of the residue $F(t)$ on $t$ in the form
\begin{equation}
  F(t) = B_{\mathrm{res}}
{\left[\frac{1}{m^2 - t}\ \exp{(R_1^2t)} + C\, \exp{(R_2^2t)} \right]}\
\label{eq:resid1}
\end{equation}
as it has been used in Refs. \cite{KaidalovP,Guaraldo} for the description of the reactions $pp \to d \pi^+$ and
$\bar p d \to p \pi^-$ at $-t \leq 1.6$ GeV$^2$.
In Eq. (\ref{eq:resid1}) the first term in the square brackets contains the nucleon pole and the second term
($\sim C$) accounts for the contribution of non-nucleonic degrees of freedom in the deuteron.

The amplitude defined by Eq.~(\ref{eq:sigt}) has a rather simple covariant structure and can be extrapolated to
large angles.
As shown in Ref. \cite{Grishina} the energy behavior of the cross section crucially depends on the form  of
the Regge trajectory $\alpha _N(t)$ for large negative $t$. The form, that better
describes the data, is the
logarithmic one with nonleading contributions:
\begin{equation} \label{nonlin} \alpha_N(t)
= \alpha_N(0)- \alpha^{\prime}_{N}(0) T_{B}\ln (1 - {t}/{T_B}). \end{equation}
Here we use this trajectory with $\alpha_{N}(0)=- 0.5,  \alpha^{\prime}_{N}(0)=0.9$~GeV${}^{-2}$, and
$T_B = 1.7$~GeV${}^2$; furthermore, we take $R=A(s,t)/B(s,t) =2$ and  adopt
the following values for the parameters of
the residue $F(t)$ in Eq. (\ref{eq:resid1}):
\begin{eqnarray}
B_{\mathrm{res}}=2.05 \cdot10^{-4}\, \mathrm{kb}^{1/2}\mathrm{GeV} ,\;
C = 0.7\ \mathrm{GeV}^{-2}  , \nonumber\\
\ R_1^2 = 2\ \mathrm{GeV}^{-2} , \;
R_2^2 = 0.03\  \mathrm{GeV}^{-2}.\; \label{Set_par}
\end{eqnarray}
Note that these parameters, except for the overall normalization factor $B_{\mathrm{res}}$, are not very different
from those determined by fitting data on the reaction $pp \to \pi^+ d$ at $-t \leq 1.6\ \mathrm{GeV}^2$
\cite{KaidalovP}.  In our case, $C$ remains unchanged and the
factor $B_{\mathrm{res}}$ and the radii $R_1^2$ and
$R_2^2$ have been fixed using two experimental values of the deuteron
photodisintegration cross section at
1.6~GeV and $\theta_p^{\mathrm{CM}}=36^{\circ}$ and $52^{\circ}$. These parameters are the same as in Ref.
\cite{Grishina}, apart from a small (about 13\%) readjustment of
$B_{\mathrm{res}}$ ($2.05 \cdot10^{-4}\, \mathrm{kb}^{1/2}$~GeV, instead of
$1.8 \cdot10^{-4}\, \mathrm{kb}^{1/2}$~GeV). This is due to the fact that in the present work the energy
dependence of the differential cross section has been calculated taking into account two amplitudes that describe
the contribution of isovector ($\rho$ like) and isoscalar ($\omega$ like) photons  (see Eq. (\ref{eq:sigt})),
while in Ref.~\cite{Grishina}  the isovector photon dominance (i.e.~$B^{\omega}=0$) was assumed.

Adopting the Vector Meson Dominance (VMD)  model we get
\begin{equation}
B^{\omega}(s,t)=B^{\rho}(s,t)/\sqrt{8}, \hspace{0.4cm}
B^{\omega}(s,u)=B^{\rho}(s,u)/\sqrt{8}.
\end{equation}

\section{Angular distributions  and forward-backward angle asymmetry}
As argued in Ref. \cite{Grishina} a forward-backward angle asymmetry in the angular distribution in the reaction
$\gamma d \to p n$ arises from the interference of the isovector and isoscalar amplitudes.
\begin{figure}[h]
\begin{center}
\leavevmode \psfig{file=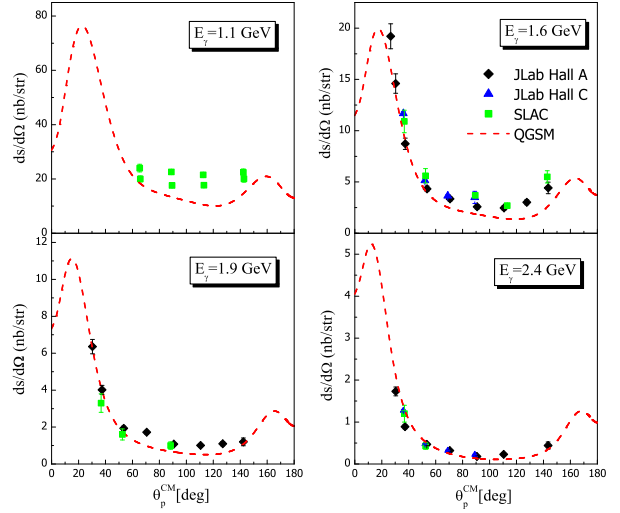,width=9.0cm}
      \caption{The differential cross section for the reaction $\gamma d \to pn$  as a function of the proton
       angle $\theta_p^{\mathrm{CM}}$ for different photon energies $E_{\gamma}$ from the QGSM (dashed lines).
       The experimental data are from SLAC \cite{Napolitano,Freedman,Beltz}, Jlab-Hall C \cite{Bochna} and
       Jlab-Hall A \cite{Gilman}.}
\label{fig:dsdo}
\end{center}
\end{figure}
\noindent In Fig.~\ref{fig:dsdo} we present    the angular dependence of ${d\sigma}/{d\Omega}$
(dashed lines) at
four photon energies $E_{\gamma}$ = 1.1 GeV, 1.6 GeV, 1.9 GeV, and 2.4 GeV. The QGSM
calculations are found to be in very good
agreement with the experimental data  taken from Refs.
\cite{Bochna,Napolitano,Freedman,Beltz,Gilman} and also with the new preliminary data from Jlab-Hall B, (not
shown in the figure), which determine almost the full angular distributions
\cite{Mirazita,Ronchetti,Rossi}.
These latter data, in particular, clearly support the predicted forward-backward angle asymmetry (see Fig. 3 of
Ref.\cite{Mirazita} and Fig. 2 of Ref. \cite{Ronchetti}). The calculated angular distributions have a dip for
$\theta_p^{\mathrm{CM}} =0^{\circ}$  and 180$^{\circ}$ which is related to the choice of the ratio
$R=A(s,t)/B(s,t)=2$. This dip does not appear for $R=1$, which  corresponds to the
limit of the Reggeized Nucleon Born Term
Model (cf. Section~2).  As argued in Ref. \cite{Mirazita} a full analysis of the data from
Jlab-Hall B will allow to
prove/disprove the appearance of the dips predicted by the QGSM.

\section{Energy dependence of the differential cross section}
\begin{figure}[h]
  \begin{center}
    \leavevmode \psfig{file=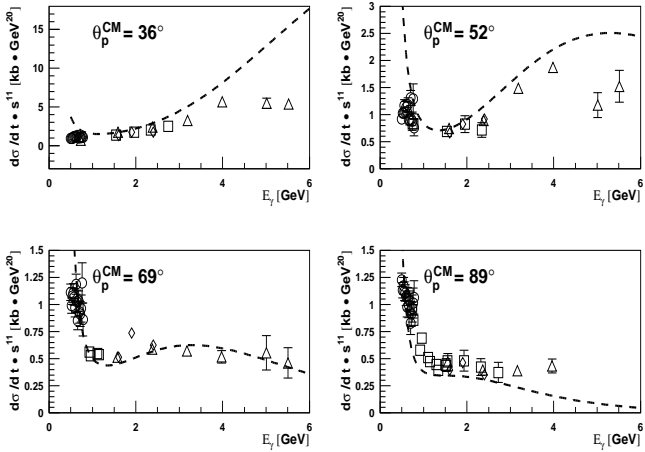,width=8.5cm,height=6.cm}
    \caption{The differential cross section for the reaction $\gamma d\to pn$ (multiplied
      by $s^{11}$) as a function of the photon lab. energy $E_{\gamma}$ at different proton angles
      $\theta_p^{\mathrm{CM}}$ in the center-of-mass frame in comparison to the experimental data from
      Mainz \cite{Crawford}, SLAC \cite{Napolitano,Freedman,Beltz}, Jlab-Hall C \cite{Bochna,Schulte}
      and Jlab-Hall A \cite{Gilman}. The dashed lines are calculated
      within the QGSM using the logarithmic nucleon
      Regge trajectory (8).}
 \label{fig:dsts11}
 \end{center}
\end{figure}

The QGSM predicts, furthermore, that the differential cross section
$d\sigma/dt$ at fixed $\theta_p^{\mathrm{CM}}$ angles decreases faster than any finite
power of $s$ and that at sufficiently large energies the perturbative regime
will become dominant. Moreover, as it was
shown in Ref. \cite{Grishina}, the model well describes the energy dependence of ${d\sigma}/{dt}\cdot s^{11}$ at
different $\theta_p^{\mathrm{CM}}$ angles for photon energies of $(1\div 4)$~GeV.
This is shown in  Fig.\ref{fig:dsts11} where the QGSM predictions for the energy dependence of
${d\sigma}/{dt}\cdot s^{11}$ at four $\theta_p^{\mathrm{CM}}$ angles -
calculated for the logarithmic nonlinear
trajectory (dashed lines) - are compared with all data available at high energies:
Mainz \cite{Crawford}, SLAC
\cite{Napolitano,Freedman,Beltz}, Jlab-Hall A \cite{Gilman}
and Jlab-Hall C \cite{Bochna,Schulte}. One can see
that above 4~GeV the QGSM overestimates the data at $\theta_p^{\mathrm{CM}}$ = 36$^{\circ}$ and systematically
underpredicts the data at~$\theta_p^{\mathrm{CM}} = 89^{\circ}$. These discrepancies
might be attributed to the simplifying assumption that all the
intermediate quark clusters have minimal spins. Moreover, the
ratio $R=A/B$ may also deviate from a constant at large momentum
transfer $t$. For a better understanding of the situation
new data at intermediate angles appear to be important.

\section{Polarization observables}

\indent
The energy dependence of the photodisintegration cross section has been shown to
be a potentially misleading indicator
for the success of pQCD. Models with asymptotic behaviour, which differ from pQCD,
 fit the data as well or even better
than pQCD (see e.g. Refs. \cite{Grishina,Brodsky,Frankfurt,Dieperink,HarryLee}). Thus
further theoretical
developments and experimental tests of nonperturbative quark models will be necessary.
 To this aim, polarization
observables are very important to further constrain the different approaches.

For the definitions of these observables in terms of helicity amplitudes we refer the reader to Ref.
\cite{Barannik}. We briefly recall here the necessary notations and definitions:
\begin{eqnarray}
F_{i,\pm}= {\langle \lambda_{p};\lambda_{n} |
\hat{T}\left(s, t\right)|
\lambda_{\gamma};\lambda_{d}\rangle },
\end{eqnarray}
where
\begin{eqnarray}
F_{1,\pm}= {\langle \pm \frac{1}{2};\pm \frac{1}{2} |
\hat{T}\left(s, t\right)|1;1\rangle },
\end{eqnarray}
\begin{eqnarray}
F_{2,\pm}= {\langle \pm \frac{1}{2};\pm \frac{1}{2}|
\hat{T}\left(s, t\right)|1;0\rangle },
\end{eqnarray}
\begin{eqnarray}
F_{3,\pm}= {\langle \pm \frac{1}{2};\pm \frac{1}{2} |
\hat{T}\left(s, t\right)|1;-1\rangle },
\end{eqnarray}
\begin{eqnarray}
F_{4,\pm}= {\langle \pm \frac{1}{2};\mp \frac{1}{2}|
\hat{T}\left(s, t\right)|1;1\rangle },
\end{eqnarray}
\begin{eqnarray}
F_{5,\pm}= {\langle \pm \frac{1}{2};\mp \frac{1}{2} |
\hat{T}\left(s, t\right)|1;0\rangle },
\end{eqnarray}
\begin{eqnarray}
F_{6,\pm}= {\langle \pm \frac{1}{2};\mp \frac{1}{2}|
\hat{T}\left(s, t\right)|1;-1\rangle }.
\label{helicityam}
\end{eqnarray}
The angular distribution in terms of the helicity amplitudes then is given by
\begin{eqnarray}
f(\theta)= \sum _{i=1}^6  \sum _{\pm} \left| F_{i,\pm}\right|^2 ,
\end{eqnarray}
while the polarization observables, i.e. the induced polarization, $P_y$, the cross section asymmetry, $\Sigma$, and the
polarization transfers, $C_{x^{\prime}}$, $C_{z^{\prime}}$ are defined via:
\begin{eqnarray}
f(\theta)\ P_y=2\ \mathrm{Im}  \sum _{i=1}^3 \left[ F_{i,+}^*
F_{i+3,-} - F_{i,-}^* F_{i+3,+}\right],
\end{eqnarray}
\begin{eqnarray}
\label{asigma}
f(\theta) \ \Sigma =-2\ \mathrm{Re} \left[ \sum _{\pm} \left( F_{1,\pm}^*
F_{3,\mp}   - F_{4,\pm} F_{6,\mp}^*\right) - \right. \nonumber \\ \left.
 F_{2,+}^* F_{2,-}+ F_{5,+}^* F_{5,-}
 \vphantom{\sum _{\pm} \left( F_{1,\pm}^*
F_{3,\mp}   - F_{4,\pm} F_{6,\mp}^*\right)}
\right],
\end{eqnarray}
\begin{eqnarray}
f(\theta)\ C_{x^{\prime}}=2\ \mathrm{Re}\sum _{\pm} \left[ F_{1,\pm}
F_{4,\mp}^*+ F_{2,\pm} F_{5,\mp}^*\right. + \nonumber \\
\left.F_{3,\pm} F^*_{6,\mp} \right],
\end{eqnarray}
\begin{eqnarray}
\label{csp}
f(\theta)\ C_{z^{\prime}} = \sum _{i=1}^6  \sum _{\pm}\left[\pm
\left| F_{i,\pm}\right|^2 \right] .
\label{polobs}
\end{eqnarray}
Note that the asymmetry $\Sigma$ is defined here with a different sign as compared to that in Ref.
\cite{Barannik} (cf. the recent review \cite{Gross} and references therein).
\begin{figure}[h]
\begin{center}
\leavevmode \psfig{file=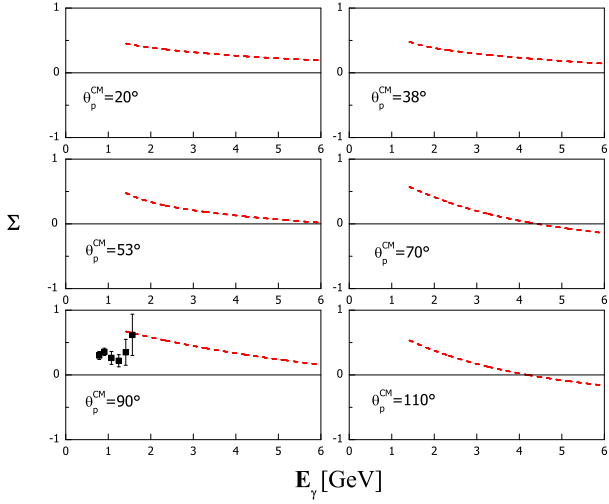,width=9cm}
\caption{The asymmetry $\Sigma$ (23) for linearly polarized photons as a function of the photon energy for different
$\theta_p^{\mathrm{CM}}$ angles. The dashed lines are calculated within the QGSM for $R=2$ while the experimental
data are taken from Ref. \cite{Adamian}. The solid line for $\Sigma = 0$ is added to guide the eye.}
\label{fig:sigmaold}
\end{center}
\end{figure}

In the following we present the QGSM predictions for the asymmetry $\Sigma$ and the polarization transfer
to the proton, $C_{z^{\prime}}$, for photon energies  $E_{\gamma} = (1.2\div 6)$~GeV.
In Fig.~\ref{fig:sigmaold} the QGSM results of the asymmetry $\Sigma$ (dashed lines) -
calculated for linearly polarized
photons - are shown as a function of the photon energy  $E_{\gamma}$
for different  angles $\theta_p^{\mathrm{CM}}$.
The QGSM predicts a slow decrease of $\Sigma (90^{\circ})$ with the photon energy from 0.6 (at 1.5 GeV) to 0.2
(at $5-6$~GeV). Also at the other angles, the asymmetry $\Sigma$ is a decreasing function of $E_{\gamma}$,
although slightly smaller in absolute magnitude. At 6~GeV it can even become negative at 70$^{\circ}$ and
110$^{\circ}$.

It is worth noticing that the behavior of the asymmetry $\Sigma$ in the QGSM is quite different from the one
predicted by the Hadron-Helicity Conservation (HHC) model as discussed in Refs. \cite{Adamian,Nagornyi}.
Moreover, according to Brodsky and Hiller \cite{Brodsky}, one should get $\lambda_d = \lambda_p + \lambda_n$
within pQCD independently of $\lambda_{\gamma}$.  Assuming that - in the scaling limit - the transverse deuteron
helicities are suppressed as compared to the longitudinal ones, the $\Sigma$ asymmetry for linearly polarized
photons
\begin{equation}
\Sigma(\theta) = (d\sigma_{||} - d\sigma_{\perp})/
(d\sigma_{||} + d\sigma_{\perp})
\end{equation}
at $\theta_p^{\rm{CM}} = 90^{\circ}$ should approach the value \cite{Nagornyi}:
\begin{equation}
\Sigma (90^{\circ}) \simeq -2\ {\mathrm{Re}}(F_{5,+}F_{5,-}^*)/ (|F_{5,+}|^2
+ |F_{5,-}|^2).
\end{equation}
Using the axial symmetry $F_{5,+}(90^{\circ})=F_{5,-}(90^{\circ})$, Nagornyi et al. \cite{Nagornyi} predicted
that $\Sigma (90^{\circ})$ should approach the value $-1$. We note, however, that the condition
$F_{5,+}(90^{\circ})= F_{5,-}(90^{\circ})$ is only valid for $\it{isoscalar}$ photons, where the isospin function
is antisymmetric. In the case of $\it{isovector}$ photons, the isospin function is symmetric and, due to the Pauli
principle, one has $F_{5,+}(90^{\circ})=-F_{5,-}(90^{\circ})$. Furthermore, according to the VMD model the isovector
photon couples to hadrons more strongly that the isoscalar photon. Thus one expects
that, in the case of hadron-helicity conservation, $\Sigma (90^{\circ})$ should not be very different from
$+1$ and, therefore,  be significantly larger than the value predicted by the QGSM.

Also shown in  Fig.~\ref{fig:sigmaold} are the experimental data from Ref. \cite{Adamian}
that are available only at
$\theta_p^{\rm{CM}} = 90^{\circ}$. The data are compatible with the QGSM predictions at 1.5 GeV:
unfortunately at lower energies, where resonance amplitudes are important, the QGSM, as well as pQCD and related
high-energy approaches, cannot be applied. Thus, polarization measurements at higher energies  are necessary to
discriminate between the models in a more adequate way.

\begin{figure}[h]
\begin{center}

\leavevmode \psfig{file=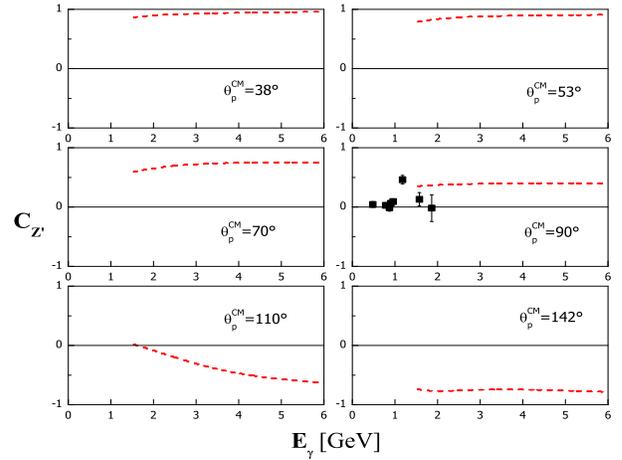,width=9cm,height=7cm}
\caption{ Polarization transfer $C_{z^{\prime}}$ for circularly polarized photons as
a function of  $E_{\gamma}$
for different angles $\theta_p^{\mathrm{CM}}$. The dashed lines are calculated within the QGSM for $R=2$ while the
experimental data are taken from  Ref. \cite{Wijesooriya} and have been corrected for spin rotation due to the
lab-CM. transformation. The  solid line for $C_{z^{\prime}}$ = 0 is added to guide the eye.}
\end{center}
\label{fig:czold}
\end{figure}

In Fig.~\ref{fig:czold} the  QGSM predictions of the polarization transfer $C_{z^{\prime}}$ for circularly polarized
photons are shown as a function of $E_{\gamma}$ at different $\theta_p^{\mathrm{CM}}$ angles. It is interesting to note
that the values of $C_{z^{\prime}}$ from the QGSM are quite large and
at $\theta_p^{\mathrm{CM}}=38^{\circ}$ almost
reach the maximal value of $\sim$1 above 2~GeV photon energy. This is directly related to the spin structure of the
amplitude defined in Eq.~(\ref{spin1}).

Also shown in Fig.~\ref{fig:czold} are the experimental data available only at
$\theta_p^{\mathrm{CM}}=90^{\circ}$
\cite{Wijesooriya}; the latter have been corrected for spin rotation due to the lab-CM transformation.
For photon energies $E_{\gamma} \geq 1.5$ GeV the data are in reasonable agreement with the QGSM
results\footnote{In our previous calculations in  Ref. \cite{GrishinaQNP}
 at $\theta_p^{\rm{CM}}=90^{\circ}$
there was an error in the sign of $C_{z^{\prime}}$ which is corrected now.}. Again, at lower energies resonant
contributions to the amplitudes are expected such that a comparison should only be meaningful for photon
energies above about 2~GeV.

The difference in the absolute value of $C_{z^{\prime}}$ between the present values and those given in
Ref. \cite{GrishinaQNP} is related to the effect of the forward-backward angle
asymmetry of the amplitude, which was not
taken into account there for polarization observables. As we have learned now, this effect is quite important not
only for the angular distribution of the differential cross section (cf. Fig. \ref{fig:dsdo}) but also for
polarization observables.

We finally  note, that the analysis of the asymmetries $C_{x^{\prime}}$ and $P_y$ is much more involved because
it is sensitive to  the re\-lative phase of the helicity amplitudes (which might also depend on the final-state
interaction of the np system). In this respect calculations of $C_{z^{\prime}}$ and $\Sigma$ are more stable
because they do not depend on this phase, but only (for $C_{z^{\prime}}$) or mainly (for $\Sigma$) on the moduli
squared of the helicity amplitudes.

\section{Summary}
The deuteron photodisintegration has been studied within the Quark-Gluon String Model, employing a logarithmic
nucleon Regge trajectory (8). The  angular distributions obtained  have been compared
to the data available, which nicely confirm
the forward-back\-ward angle asymmetry predicted by the model.

In addition, new results from the QGSM for the polarization transfer to the proton, $C_{z^{\prime}}$ and the cross
section asymmetry, $\Sigma$, for photon energies $(1.2\div 6)$~GeV and at different proton CM-angles,
$\theta_p^{\rm{CM}}$, have been calculated. The results have been compared to the data available only at
$\theta_p^{\rm{CM}}$ = 90$^{\circ}$ and up to 2~GeV; for photon energies
$\geq 1.5$~GeV the data are found in
reasonable agreement with the QGSM results. Since contributions from
resonant amplitudes should be present in the data, a meaningful comparison with the QGSM results can only be
performed for higher energies. Data at high energy should come up in the near future from Jlab \cite{Ron} and
will allow to discriminate between the different approaches discussed in this work.

\vspace{0.5cm} We are grateful to Ronald Gilman for useful comments and for sending us the experimental data on
the polarization transfer $C_{z^{\prime}}$ in Fig.~\ref{fig:czold} corrected for spin rotation due to the lab-CM
transformation. The work has been supported by the Italian Istituto Nazionale di Fisica Nucleare (INFN), the Russian
Fund for Basic Research (grant 02-02-16783) and by the Federal Program of the Russian Ministry of Industry,
Science and Technology No. 40.052.1.1.1112.

\end{document}